\documentclass[aps,prl,twocolumn,superscriptaddress, amsmath,amssymb]{revtex4-2}

\usepackage{diagbox}
\usepackage{ragged2e}
\usepackage{array}
\usepackage{booktabs}
\usepackage{graphicx}
\usepackage{dcolumn}
\usepackage{bm}
\usepackage[colorlinks, linkcolor=black, anchorcolor=black, citecolor=black]{hyperref}

\begin{document}

\preprint{APS/123-QED}

\title{Momentum-resolved thermodynamics and barrier-free order-parameter collapse in spin-orbit-coupled superconductors under a Zeeman field}

\author{Xusheng Wang}
\email{wang-xs23@mails.tsinghua.edu.cn}
\affiliation{%
State Key Laboratory of Low-Dimensional Quantum Physics, Department of Physics, Tsinghua University, Beijing 100084, China}
\author{Gaomin Tang}
\affiliation{
Graduate School of China Academy of Engineering Physics, Beijing 100193, China}
\author{Lianyi He}
\affiliation{%
State Key Laboratory of Low-Dimensional Quantum Physics, Department of Physics, Tsinghua University, Beijing 100084, China}
\author{Shuai-Hua Ji}
\affiliation{%
State Key Laboratory of Low-Dimensional Quantum Physics, Department of Physics, Tsinghua University, Beijing 100084, China}
\affiliation{%
Frontier Science Center for Quantum Information, Beijing 100084, China}

\date{\today}

\begin{abstract}

Spin-orbit coupling (SOC) can enhance the critical Zeeman field of clean superconductors beyond the Pauli limit. We develop a momentum decomposition of the zero-temperature free energy to investigate spin-singlet $s$-wave superconductors with general time-reversal-symmetric SOC. The SOC component perpendicular to the Zeeman field enhances the critical field and suppresses first-order features, whereas the parallel component favors Pauli-limited thermodynamic behavior. Applying this framework to Rashba superconductors and Ising superconductors with weak Rashba SOC, we identify the origin of the abrupt collapse of the superconducting order parameter at the critical field. Competing angular contributions produce a nearly flat free-energy landscape, allowing a barrier-free order-parameter collapse distinct from the conventional Pauli-limited first-order transition.

\end{abstract}

\maketitle

In ultra-thin superconductors, the reduced thickness strongly suppresses orbital depairing by in-plane magnetic fields, making the Zeeman effect the dominant pair-breaking mechanism \cite{fulde1973high}. For a conventional spin-singlet $s$-wave superconductor with spatially uniform pairing, the zero-temperature paramagnetic critical field is the Pauli limit $B_p$ \cite{clogston1962upper,chandrasekhar1962note}, at which the superconducting state undergoes a first-order transition (FOT) to the normal state \cite{sarma1963influence,maki1966effect}.

Within the spin-singlet pairing framework, spin-orbit interactions provide important mechanisms for exceeding the Pauli limit. In superconductors with sizable impurities, spin-orbit scattering mixes spin states and reduces paramagnetic depairing, thereby enhancing the critical field \cite{klemm1975theory,tedrow1982critical}. In clean and non-magnetic systems, intrinsic SOC that preserves time-reversal symmetry can instead protect superconductivity through spin-momentum locking \cite{ghosh2021recent,amundsen2024colloquium}. Rashba and Ising SOC, associated with out-of-plane and in-plane inversion asymmetries, are two prominent forms of SOC in two-dimensional superconductors \cite{uchihashi2021surface}. For spatially uniform pairing in the conventional Rashba model, the critical-field enhancement is limited to $\sqrt{2}B_p$ \cite{gor2001superconducting,barzykin2002inhomogeneous,olde2021tunable,liu2018interface,yoshizawa2021atomic}. By contrast, Ising SOC can support substantially larger in-plane critical fields \cite{yuan2016ising,liu2017unconventional,ilic2017enhancement,mockli2020ising,liu2020microscopic,zhou2016ising}.

Ising superconductivity has been extensively investigated in monolayer transition-metal dichalcogenides \cite{zhu2011giant,lu2015evidence,saito2016superconductivity,xi2016ising,sohn2018unusual,yi2022crossover,de2018tuning,cho2022nodal,li2021recent,wang2021ising,wan2023orbital,ji2024continuous,volavka2026ising}. In realistic devices, interfacial interactions and electrostatic gating can additionally generate Rashba SOC \cite{lu2015evidence,saito2016superconductivity,wang2015strong,masseroni2024spin}. Although typically weaker than the Ising component, Rashba SOC can be comparable to the superconducting energy scale and substantially modify the magnetic response. Previous theoretical studies have examined its influence on the temperature dependence of the critical field \cite{lu2015evidence,saito2016superconductivity}. More recently, an abrupt collapse of the superconducting order parameter was predicted at low temperatures in an Ising superconductor with Rashba SOC \cite{harms2025collapse}. This behavior raises a thermodynamic question: does the collapse mark an equilibrium FOT or the disappearance of a metastable superconducting solution? Resolving this distinction requires an explicit free-energy analysis.

In this Letter, we develop a momentum decomposition of the zero-temperature free energy to investigate spin-singlet $s$-wave superconductors with general time-reversal-symmetric SOC. We show that the perpendicular SOC component enhances the critical field and suppresses first-order features, whereas the parallel component restores Pauli-like thermodynamic behavior in the strong-SOC regime. Applying this framework to Rashba superconductors and Ising superconductors with Rashba SOC, we reveal that the abrupt collapse of the superconducting order parameter arises from the cancellation of competing angular contributions. The resulting nearly flat critical-field free-energy landscape distinguishes this barrier-free order-parameter collapse from the conventional Pauli-limited first-order transition.

We consider a spin-singlet $s$-wave superconductor with general time-reversal-symmetric SOC subject to a Zeeman field. The normal-state Hamiltonian reads \cite{harms2025collapse,wang2026universal}
\begin{equation}\label{normal_state}
    \hat{H}_N(\mathbf{p})=(\xi_\mathbf{p}-\mu)\sigma_0+\left[\mathbf{g}(\mathbf{p})-\mathbf{H}\right]\cdot\boldsymbol{\sigma},
\end{equation}
where $\xi_\mathbf{p}$ is the kinetic energy, $\mu$ is the chemical potential, and $\sigma_0$ and $\boldsymbol{\sigma}=(\sigma_x,\sigma_y,\sigma_z)$ denote the identity matrix and the vector of Pauli matrices in spin space, respectively. The SOC is described by the momentum-dependent vector $\mathbf{g}(\mathbf{p})$, which satisfies $\mathbf{g}(-\mathbf{p})=-\mathbf{g}(\mathbf{p})$ owing to time-reversal symmetry \cite{Frigeri2004}. The Zeeman field is defined as $\mathbf{H}=g_L\mu_B\mathbf{B}/2$, where $g_L$ is the Landé $g$ factor, $\mu_B$ is the Bohr magneton, and $\mathbf{B}$ is the applied in-plane magnetic field. We choose the $x$ axis along $\mathbf{B}$.

\begin{figure*}[htbp]
\centering
\includegraphics[width=\linewidth]{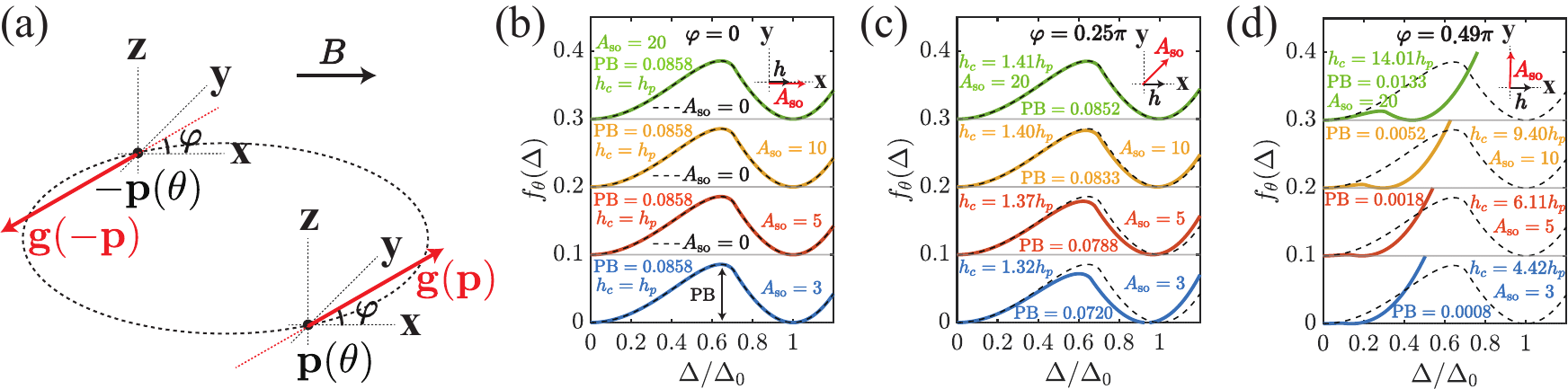}
\caption{
(a) Schematic spin texture of a general time-reversal-symmetric SOC at $\pm\mathbf{p}(\theta)$, satisfying $\mathbf{g}(-\mathbf{p})=-\mathbf{g}(\mathbf{p})$.
(b-d) Angular-resolved free-energy landscapes for different SOC strengths at (b) $\varphi=0$, (c) $\varphi=0.25\pi$, and (d) $\varphi=0.49\pi$. In (d), portions with $f_\theta>0.1$ are omitted. Here, $h_p$ denotes the Pauli-limited Zeeman field. Black dashed lines show the corresponding Pauli-limited free-energy landscapes for reference. The curves are rescaled and vertically offset by increments of $0.1$ for clarity.
}
\label{general_illustration}
\end{figure*}

In the Nambu basis $(\hat{c}_{\mathbf{p},\uparrow}, \hat{c}_{\mathbf{p},\downarrow}, \hat{c}_{-\mathbf{p},\uparrow}^\dagger, \hat{c}_{-\mathbf{p},\downarrow}^\dagger)^{\mathrm{T}}$, the Bogoliubov--de Gennes Hamiltonian takes the form \cite{patil2023spectral,wang2025unified,wang2026first}:
\begin{equation}\label{BdG_eq}
    \hat{H}_{\mathrm{BdG}}(\mathbf{p})=    
    \begin{pmatrix}
    \hat{H}_N(\mathbf{p}) & i\Delta \sigma_y \\
    -i\Delta \sigma_y & -\hat{H}^*_N(-\mathbf{p})
    \end{pmatrix},
\end{equation}
where $\Delta$ is the spatially uniform superconducting order parameter, chosen to be real by a global gauge transformation. Diagonalizing Eq.~(\ref{BdG_eq}) yields four eigenvalues $E_{\mathbf{p},i}$, with $i=1,2,3,4$. Up to terms independent of $\Delta$, the superconducting free energy $F$ is \cite{altland2010condensed,xie2020spin,li2024topological,wang2025temperature}
\begin{equation}\label{free_energy_finite_T}
    F=\frac{\Delta^2}{g}+\frac{1}{2\beta}\sum_{\mathbf{p}}\ln (1+e^{-\beta E_{\mathbf{p},i}}),
\end{equation}
where $g$ is the superconducting coupling strength and $\beta=1/(k_BT)$. The equilibrium order parameter is determined by the global minimum of $F$ with respect to $\Delta$. Although Eq.~(\ref{free_energy_finite_T}) applies at arbitrary temperature, we focus on the zero-temperature limit to elucidate the low-temperature thermodynamics.

In two dimensions, the momentum direction is parameterized by a single azimuthal angle $\theta$, measured from the $x$ axis along the applied Zeeman field. The quasiparticle energies can therefore be written as $E_{\mathbf{p},i}=E_{\mathrm{p},i}(\theta)$, where $\mathrm{p}=|\mathbf{p}|$. Assuming that the Fermi energy greatly exceeds the superconducting, SOC, and Zeeman energy scales, we evaluate the SOC on the Fermi surface and neglect its dependence on the radial energy variable:
$\mathbf{g}(\mathbf{p})\simeq\mathbf{g}[\mathbf{p}(\theta)]\equiv\mathbf{g}(\theta)$. The zero-temperature free energy becomes \cite{supp}
\begin{equation}\label{zero_T_free_energy}
    F = \frac{1}{2\pi}\int_0^{2\pi}\left[\frac{\Delta^2}{g }-N\int_0^{E_c}\sum_{i=1}^4\frac{\bigl|E_{\mathrm{p},i}(\theta)\bigr|}{2} \mathrm{d}\epsilon_{\mathrm{p}}\right]\mathrm{d}\theta,
\end{equation}
where $N$ is the normal-state density of states at the Fermi level and $E_c$ is the energy cutoff. Equation (\ref{zero_T_free_energy}) expresses $F$ as an angular average of
\begin{equation}\label{sliced_free_energy}
    F_{\theta}=\frac{\Delta^2}{g }-N\int_0^{E_c}\sum_{i=1}^4\frac{\bigl|E_{\mathrm{p},i}(\theta)\bigr|}{2} \mathrm{d}E_{\mathrm{p}}.
\end{equation}
Although the equilibrium properties are determined by the total free energy, $F_\theta$ reveals how individual momentum directions contribute to the superconducting response.

According to Eqs. (\ref{normal_state}) and (\ref{BdG_eq}), the quasiparticle spectra $E_{\mathrm{p},i}$ are fully characterized by three parameters \cite{supp}: $|\mathbf{g}(\theta)|$, $|\mathbf{H}|$, and the angle $\varphi$ between the SOC vector and the Zeeman field [Fig.~\ref{general_illustration}(a)]. Symmetry ensures that the orientations $\varphi_0$, $\pi-\varphi_0$, and $\pi+\varphi_0$ yield identical contributions to $F_\theta$. We therefore restrict $\varphi$ to $[0,\pi/2]$, defining it as the smaller angle between the axes specified by the two vectors. Examining $F_\theta$ as a function of these three parameters provides insight into how SOC influences superconductivity under a Zeeman field.

For convenience, we introduce the dimensionless quantities
\begin{equation}
    h= \frac{|\mathbf{H}|}{\Delta_0},\  A_{\mathrm{so}}=\frac{|\mathbf{g}(\theta)|}{\Delta_0},\ f_\theta(\Delta)= \frac{F_{\theta}(\Delta)-F_{\theta}(0)}{N\Delta_0^2},
\end{equation}
where $\Delta_0$ is the zero-temperature superconducting order parameter in the absence of a Zeeman field.

We first examine the angular-resolved free-energy landscapes for different SOC orientations $\varphi$ and strengths $A_{\mathrm{so}}$, as shown in Figs.~\ref{general_illustration}(b)-\ref{general_illustration}(d). For each angular contribution, the critical field is identified as the field at which $\Delta=0$ first becomes a global minimum of $F_\theta$. The free-energy landscape at this field also distinguishes the transition order: an FOT exhibits degenerate superconducting and normal-state minima separated by a potential barrier, whereas a second-order transition occurs as the superconducting minimum continuously approaches $\Delta=0$ \cite{wang2026first}.

To illustrate the SOC effects, we include the Pauli-limited landscape ($A_{\mathrm{so}}=0$) as a reference for each curve. When the SOC vector is parallel to the Zeeman field ($\varphi=0$), both the critical field and the free-energy landscape remain nearly identical to their Pauli-limited counterparts [Fig.~\ref{general_illustration}(b)]. This behavior follows from the quasiparticle dispersions \cite{wang2026universal}
\begin{equation}\label{H_para}
    E_{\mathbf{p}}=\pm \sqrt{\Delta^2+(\xi_{\mathbf{p}}+s |\mathbf{g}(\theta)|)^2}+ s\mathrm{H},\ s=\pm 1.
\end{equation}
Introducing $\tilde{\xi}_{\mathbf{p},s}=\xi_{\mathbf{p}}+s|\mathbf{g}(\theta)|$ reduces these dispersions to the Pauli-limited form \cite{meservey1970magnetic,meservey1975tunneling,meservey1994spin,gubbels2013imbalanced}. For $E_c\gg|\mathbf{g}(\theta)|$, the resulting shifts of the integration boundaries have a negligible effect on the superconducting free-energy difference in Eq.~(\ref{sliced_free_energy}).

When the SOC vector has both parallel and perpendicular components ($\varphi=0.25\pi$), the thermodynamic behavior deviates from the Pauli-limited case [Fig.~\ref{general_illustration}(c)]. The critical field $h_c$ increases monotonically with $A_{\mathrm{so}}$, whereas the potential-barrier height $\mathrm{PB}$ initially decreases and subsequently increases. This behavior suggests that the perpendicular component dominates the suppression of first-order features at moderate SOC strengths ($A_{\mathrm{so}}<5$), while the parallel component becomes dominant in the strong-SOC regime, restoring Pauli-like first-order behavior. At $A_{\mathrm{so}}=20$, the critical-field free-energy landscape closely resembles the Pauli-limited reference.

For a purely perpendicular SOC vector ($\varphi=\pi/2$), the idealized Ising superconductor exhibits a divergent critical field at zero temperature \cite{ilic2017enhancement}. We therefore consider the nearly perpendicular configuration $\varphi=0.49\pi$ [Fig.~\ref{general_illustration}(d)]. Although the first-order features are substantially weaker in this configuration, $h_c$ and $\mathrm{PB}$ exhibit the same qualitative dependence on SOC strength as at $\varphi=0.25\pi$.

The effect of SOC orientation can also be examined at fixed $A_{\mathrm{so}}$. As $\varphi$ increases from $0$ to $0.49\pi$, the critical field rises and the first-order features weaken [Figs.~\ref{general_illustration}(b)-\ref{general_illustration}(d)]. These results indicate that the perpendicular SOC component enhances resistance against Zeeman depairing while suppressing the first-order character of the transition.

\begin{figure}[ht]
\centering
\includegraphics[width=0.85\linewidth]{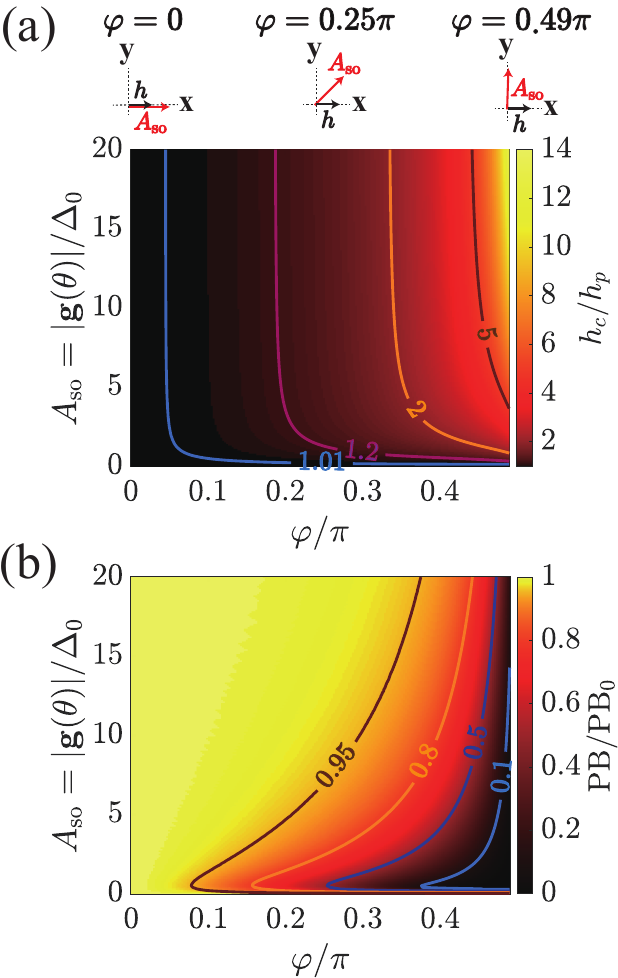}
\caption{
(a) Critical field $h_c$ and (b) potential-barrier height $\mathrm{PB}$ as functions of the SOC orientation $\varphi$ and strength $A_{\mathrm{so}}$. The color scales represent $h_c/h_p$ and $\mathrm{PB}/\mathrm{PB}_0$, respectively, where $h_p$ and $\mathrm{PB}_0$ are the corresponding Pauli-limited values. The barrier height is evaluated at the corresponding critical field.
}
\label{hc_PB_distribution}
\end{figure}

To further elucidate the effects of SOC orientation and strength, we map $h_c(\varphi,A_{\mathrm{so}})$ and $\mathrm{PB}(\varphi,A_{\mathrm{so}})$ in Fig.~\ref{hc_PB_distribution}. The contours in Fig.~\ref{hc_PB_distribution}(a) show that $h_c$ increases with both $A_{\mathrm{so}}$ and $\varphi$, consistent with the trends discussed above. Figure~\ref{hc_PB_distribution}(b) confirms that $\mathrm{PB}$ varies non-monotonically with SOC strength but decreases monotonically as the SOC vector rotates toward the perpendicular orientation. According to Eqs.~(\ref{zero_T_free_energy}) and (\ref{sliced_free_energy}), the total free energy for a general time-reversal-symmetric SOC texture is obtained by integrating the angular-resolved contributions with the appropriate weights. Analyzing representative $\varphi$ slices thus provides a useful framework for understanding the superconducting response to a Zeeman field in systems with more complex SOC textures.

\begin{figure*}[htbp]
\centering
\includegraphics[width=1.0\linewidth]{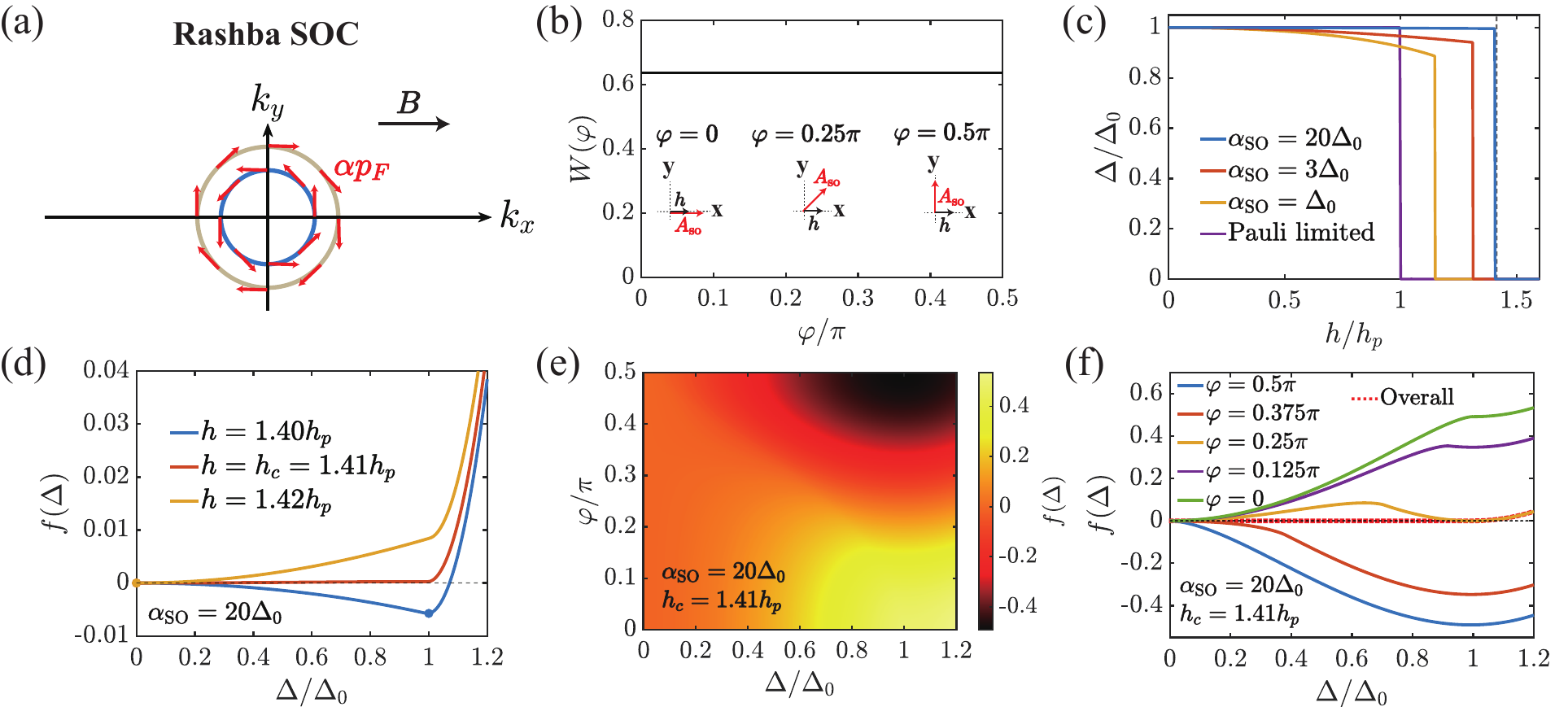}
\caption{
Collapse of the superconducting order parameter and its origin in Rashba superconductors.
(a) Schematic Rashba spin texture.
(b) Normalized angular weight $W(\varphi)$ of the $\varphi$ slices.
(c) Zeeman-field dependence of the superconducting order parameter for different Rashba SOC strengths.
(d) Free-energy landscapes near the critical field for $\alpha_{\mathrm{so}}=20\Delta_0$ ($A_{\mathrm{SO}}=20$). The black dashed line corresponds to $h=\sqrt{2}h_p$.
(e) Angular-resolved free-energy landscapes at the critical field for $\alpha_{\mathrm{SO}}=20\Delta_0$. The color scale represents the free energy.
(f) Representative $\varphi$-resolved free-energy profiles from (e), together with the total free-energy profile.
}
\label{Rashba_SOC}
\end{figure*}

We first consider Rashba superconductors, described by $\mathbf{g}(\mathbf{p})=(-\alpha_{R} p_y, \alpha_{R} p_x,0)$ \cite{gor2001superconducting}, where $\alpha_R$ is the Rashba SOC coefficient. For a circular Fermi surface with radius $p_F$, the SOC vector becomes
\begin{equation}\label{Rashba_SOC_form}
    \mathbf{g}(\theta)=\alpha_{\mathrm{so}}(-\sin \theta, \cos \theta,0),
\end{equation}
where $\alpha_{\mathrm{SO}}=\alpha_Rp_F$ is the momentum-independent SOC strength. The corresponding spin texture and normalized angular weight $W(\varphi)$ are shown in Figs.~\ref{Rashba_SOC}(a) and \ref{Rashba_SOC}(b), respectively. The uniform distribution of $\varphi$ over $[0,\pi/2]$ implies that the total free energy is an equally weighted average of the $\varphi$-resolved contributions.

\begin{figure*}[htbp]
\centering
\includegraphics[width=1.0\linewidth]{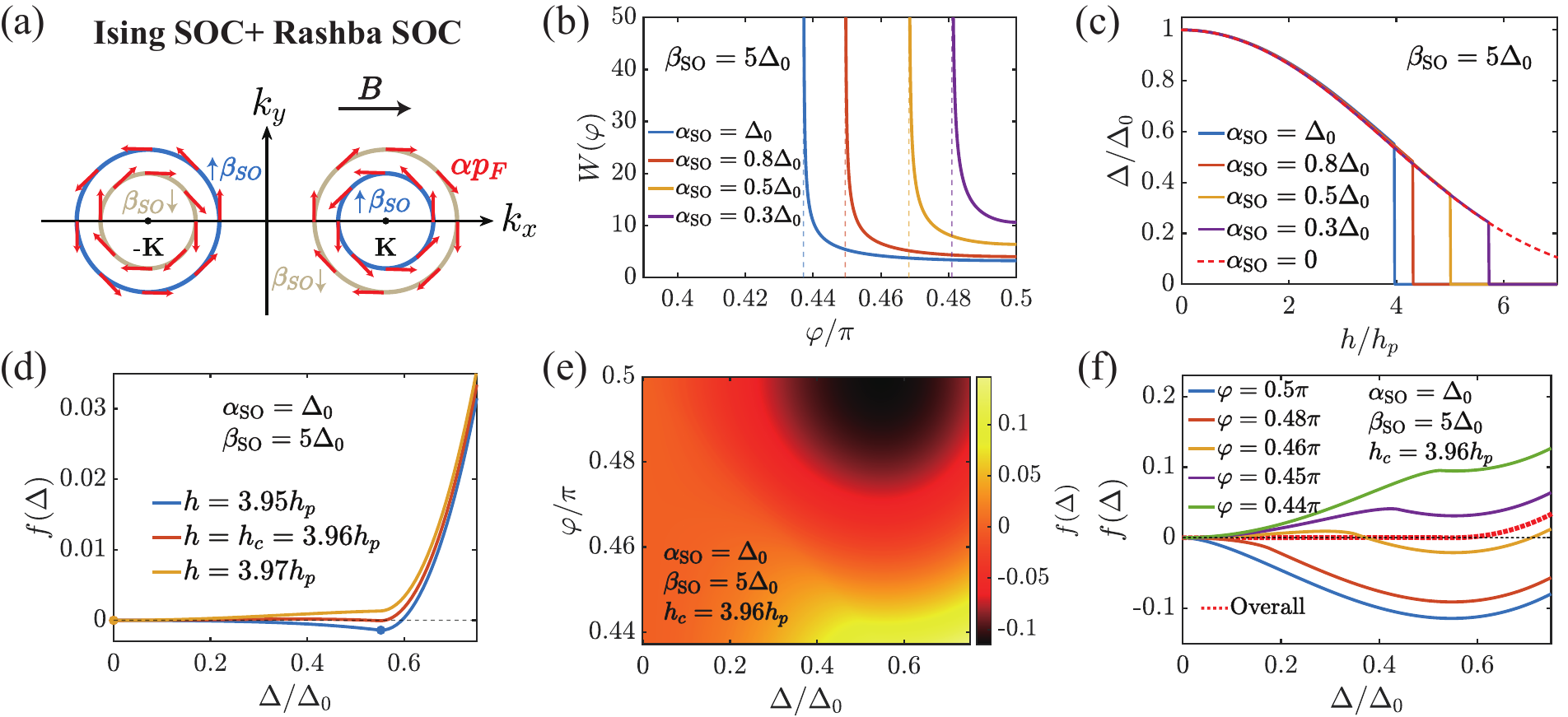}
\caption{
Collapse of the superconducting order parameter and its origin in Ising superconductors with weak Rashba SOC.
(a) Schematic spin texture of coexisting Ising and Rashba SOC.
(b) Normalized angular weight $W(\varphi)$ of the $\varphi$ slices for different Rashba SOC strengths.
(c) Zeeman-field dependence of the superconducting order parameter for different Rashba SOC strengths.
(d) Free-energy landscapes near the critical field for $\alpha_{\mathrm{SO}}=\Delta_0$.
(e) Angular-resolved free-energy landscapes at the critical field for $\alpha_{\mathrm{SO}}=\Delta_0$ and $\beta_{\mathrm{SO}}=5\Delta_0$. The color scale represents the free energy.
(f) Representative $\varphi$-resolved free-energy profiles from (e), together with the total free-energy profile.
}
\label{Rashba_plus_Ising_SOC}
\end{figure*}

Figure~\ref{Rashba_SOC}(c) presents the equilibrium order parameter as a function of Zeeman field for different Rashba SOC strengths. As $\alpha_{\mathrm{so}}$ increases, the critical field rises from $h_p$ toward $\sqrt{2}h_p$, consistent with previous studies \cite{gor2001superconducting}. For weak Rashba SOC, the order parameter is slightly suppressed even below $h_p$. Similar behavior occurs in Ising superconductors [Fig.~\ref{Rashba_plus_Ising_SOC}(c)] and originates from the SOC component perpendicular to the Zeeman field. Notably, both Pauli-limited and Rashba superconductors exhibit an abrupt collapse of the order parameter at their respective critical fields. Although such a discontinuity commonly accompanies an FOT, the free-energy landscapes reveal a distinct mechanism in the Rashba case.

To elucidate this distinction, Fig.~\ref{Rashba_SOC}(d) shows the free-energy landscapes near the critical field for $\alpha_{\mathrm{so}}=20\Delta_0$. Just below the critical field ($h=1.40h_p$), the free energy has a single minimum near $\Delta=\Delta_0$. Just above it ($h=1.42h_p$), the minimum shifts to $\Delta=0$. At $h_c\simeq1.41h_p$, the landscape becomes nearly flat over $\Delta\in[0,\Delta_0]$, with no intervening potential barrier. The collapse therefore differs from the conventional Pauli-limited FOT, whose critical-field landscape contains two isolated, degenerate minima separated by a barrier.

This nearly flat landscape results from the angular average of the $\varphi$-resolved contributions, as illustrated in Figs.~\ref{Rashba_SOC}(e) and \ref{Rashba_SOC}(f). Contributions with $\varphi<0.25\pi$ favor $\Delta=0$, whereas those with $\varphi>0.25\pi$ favor a finite order parameter near $\Delta_0$. Their competing contributions approximately cancel over $\Delta\in[0,\Delta_0]$, producing the nearly degenerate total free-energy landscape.

We apply the same free-energy analysis to Ising superconductors with weak Rashba SOC. The corresponding spin texture, illustrated in Fig.~\ref{Rashba_plus_Ising_SOC}(a), takes the form \cite{lu2015evidence}
\begin{equation}
    \mathbf{g}(\theta)=(-\alpha_{\mathrm{SO}}\sin \theta, \alpha_{\mathrm{SO}}\cos \theta, s\beta_{\mathrm{SO}}),
\end{equation}
where $\beta_{\mathrm{SO}}$ is the Ising SOC strength and $s=\pm1$ labels the two valleys. To examine the Rashba contribution, we vary $\alpha_{\mathrm{SO}}$ over values smaller than or comparable to $\Delta_0$, while fixing $\beta_{\mathrm{SO}}=5\Delta_0$.

Figure~\ref{Rashba_plus_Ising_SOC}(b) shows the normalized angular weights $W(\varphi)$. Although $W(\varphi)$ diverges at the minimum allowed angle for each SOC configuration, this divergence is integrable \cite{supp}. The resulting Zeeman-field-dependent equilibrium order parameters are shown in Fig.~\ref{Rashba_plus_Ising_SOC}(c). Increasing the Rashba SOC strength extends the angular distribution toward smaller $\varphi$, introducing contributions with larger SOC components parallel to the Zeeman field and thereby reducing the critical field. The $\Delta(h)$ curves also reproduce the abrupt collapse of the order parameter reported previously \cite{harms2025collapse}. Unlike the collapse from approximately $\Delta_0$ to zero in Rashba superconductors, the magnitude of the discontinuity here depends strongly on the Rashba component, with larger $\alpha_{\mathrm{SO}}$ producing a larger jump.

To examine the thermodynamic character of this collapse, we consider the free-energy landscapes for $\alpha_{\mathrm{SO}}=\Delta_0$ [Fig.~\ref{Rashba_plus_Ising_SOC}(d)]. Similar with the Rashba case, the corresponding critical-field landscape exhibits no discernible potential barrier and is nearly flat over $\Delta\in[0,0.6\Delta_0]$. This behavior also differs from the conventional double-well landscape of a Pauli-limited FOT. The near-degeneracy again results from the cancellation of competing $\varphi$-resolved contributions, as illustrated in Figs.~\ref{Rashba_plus_Ising_SOC}(e) and \ref{Rashba_plus_Ising_SOC}(f).

These results attribute the collapse to the Rashba-induced distribution of SOC orientations relative to the Zeeman field. Different $\varphi$ slices have distinct angular-resolved critical fields. At the total critical field, some contributions favor the normal state, whereas others favor a finite order parameter. Their weighted sum produces a nearly flat free-energy landscape near $\Delta=0$, allowing the equilibrium order parameter to collapse abruptly without an intervening barrier.

In summary, we have developed a momentum decomposition of the free energy that provides a framework for understanding superconductors with general momentum-dependent time-reversal-symmetric SOC. This approach resolves the effects of SOC strength and orientation relative to the Zeeman field. The perpendicular SOC component enhances the critical field and suppresses first-order features, whereas the parallel component favors Pauli-limited thermodynamic behavior. Such first-order features may manifest experimentally as magnetic-field hysteresis \cite{tedrow1977supercooling} or thermodynamic discontinuities \cite{bianchi2002first}. Applying this framework to Rashba superconductors and Ising superconductors with Rashba SOC, we identify a barrier-free collapse of the superconducting order parameter associated with a nearly flat critical-field free-energy landscape. This behavior originates from the cancellation of competing angular contributions introduced by Rashba SOC and differs from the conventional Pauli-limited FOT.

Monolayer transition-metal dichalcogenides \cite{lu2015evidence,saito2016superconductivity,xi2016ising,sohn2018unusual,yi2022crossover} and atomically thin lead \cite{liu2018interface} and indium layers \cite{yoshizawa2021atomic} provide candidate platforms for testing these predictions. Scanning tunneling spectroscopy under an in-plane magnetic field could probe the spectral signatures of the predicted collapse, while complementary thermodynamic measurements could establish the transition character.

\textit{Acknowledgments.--} X. W acknowledges stimulating discussions with K. T. Law and N. F. Yuan. This work was supported by the Quantum Science and Technology National Science and Technology Major Project (Grant No. 2023ZD0300500), and the National Natural Science Foundation of China (Grant Nos. 52388201 and 12374048).

\nocite{*}

\bibliographystyle{Cite_form}
\bibliography{bib_main}

\end{document}